\documentclass{article}

\usepackage{amsmath, amsfonts, natbib, graphicx, mathtools}
\usepackage{hyperref}

\graphicspath{{figures/}}

\hypersetup{
    colorlinks,%
    citecolor=black,%
    filecolor=black,%
    linkcolor=black,%
    urlcolor=black
}

\title{Efficient Bayesian inference in stochastic chemical kinetic models using graphical processing units}
\author{Jarad Niemi and Matthew Wheeler}
\date{\today}

\begin{document}

\maketitle

\begin{abstract}
A goal of systems biology is to understand the dynamics of intracellular systems. Stochastic chemical kinetic models are often utilized to accurately capture the stochastic nature of these systems due to low numbers of molecules. Collecting system data allows for estimation of stochastic chemical kinetic rate parameters. We describe a well-known, but typically impractical data augmentation Markov chain Monte Carlo algorithm for estimating these parameters. The impracticality is due to the use of rejection sampling for latent trajectories with fixed initial and final endpoints which can have diminutive acceptance probability. We show how graphical processing units can be efficiently utilized for parameter estimation in systems that hitherto were inestimable. For more complex systems, we show the efficiency gain over traditional CPU computing is on the order of 200. Finally, we show a Bayesian analysis of a system based on Michaelis-Menton kinetics.
\end{abstract}

\section{Introduction}

The development of highly parallelized graphical processing units (GPUs) has largely been driven by the video game industry for faster and more accurate real-time 3D visualization. More recently the graphics card industry has introduced general purpose GPUs for scientific computing. Much of the focus of this work has been on building parallelized linear algebra routines since these applications are inherently amenable to parallelization \citep{Galo:Govi:Hens:Mano:lu:2005, Volk:Demm:benc:2008, Krug:West:line:2005}. More recently the biological scientific community has taken interest for applications such as leukocyte tracking \citep{Boye:Tarj:Acto:Skad:acce:2009}, cluster identification in flow cytometry \citep{Such:Wang:Chan:Frel:Cron:West:unde:2010}, and molecular dynamic simulation of intracellular processes \citep{Li:Petz:effi:2010}.

The statistics community has been slower to venture into this massively parallel area, but in the last couple of years a few papers have appeared using GPUs to provide efficient analyses of problems that would otherwise by computationally prohibitive. Topics of these papers include statistical phylogenetics \citep{Such:Ramb:many:2009}, slice sampling \citep{Tibb:Hara:Liec:para}, high-dimensional optimization \citep{Zhou:Lang:Such:graph:2010}, simulation of Ising models \citep{Prei:Virn:Paul:Schn:gpu:2009}, approximate Bayesian computation (ABC) \citep{Liep:Barn:Cule:Ergu:Krik:Toni:Stum:abc:2010}, estimating multivariate mixtures \citep{Such:Wang:Chan:Frel:Cron:West:unde:2010}, population-based Markov chain Monte Carlo (MCMC) and sequential Monte Carlo methods \citep{Lee:Yau:Gile:Douc:Holm:on:2009}. As the number of parallel cores increase, the presence of GPU computing will undoubtably grow.

The use of parallel computing in the area of stochastic chemical kinetics is primarily focused on simulation of systems assuming known reaction parameters \citep{Li:Petz:effi:2010}. Some have used these simulations within an approximate Bayesian computation (ABC) framework for estimation of reaction parameters \citep{Liep:Barn:Cule:Ergu:Krik:Toni:Stum:abc:2010}. The Bayesian approach presented here is a special-case of the ABC methodology and can be used as a gold-standard for comparing efficacy of the more general ABC methodology. This approach implements data augmentation MCMC (DA-MCMC) where the latent trajectories for chemical species are simulated at each iteration of the MCMC algorithm \cite{Marj:Moli:Plag:Tava:mark:2003}. Coupling this algorithm with GPU computing provides Monte Carlo estimates of parameter posteriors for sizeable systems in reasonable time frames.

This article proceeds as follows. In section \ref{sec:model}, we describe stochastic chemical kinetic models. In section \ref{sec:inference}, we introduce the DA-MCMC Bayesian approach to parameter inference in these models. Section \ref{sec:gpu-adaption} discusses modifications required or beneficial for using this inferential technique on GPUs. Section \ref{sec:simulation} provides a simulation study to determine the computational efficiency gain of using GPUs relative to CPUs as well as full Bayesian analysis of a Michaelis-Menton system. Finally, concluding remarks and future research plans are discussed in section \ref{sec:discussion}.

\section{Stochastic chemical kinetic models \label{sec:model}}

Many biological phenomenon can be modeled using stochastic chemical kinetic models \cite{Wilk:stoc:2006}. These models are particularly useful when at least one species has a small number of molecules and therefore deterministic models provide poor approximations. In biology, this is common when considering intracellular populations such as the number of DNA, RNA, and protein molecules. Below we introduce the notation required for understanding these stochastic chemical kinetic models as well as methods used to simulate from them.

Consider a spatially homogeneous biochemical system within a fixed volume at constant temperature. This system contains $N$ species $\{S_1,\ldots,S_N\}$ with state vector $X(t)=(X_1(t),\ldots,X_N(t))'$ describing the number of molecules of each species at time $t$. This state vector is updated through $M$ reactions labeled $R_1,\ldots, R_M$. Reaction $j\in\{1,\ldots,M\}$ has a \emph{propensity} $a_j(x)=\theta_j h_j(x)$ where $\theta_j$ is the unknown stochastic reaction rate parameter for reaction $j$ and $h_j(x)$ is a known function of the system state $x$. Multiplying the propensity by an infinitesimal $\tau$ provides the probability of reaction $j$ occurring in the interval $[t,t+\tau)$.  If reaction $j$ fires, the state vector is updated to $X(t+\tau)=X(t)+v_j$ where $v_j=(v_{1j}, \ldots, v_{Nj})'$ describes the number of molecules of each species that are consumed or produced in reaction $j$.

The probability distribution for the state at time $t$, $p(t,x)$, is the solution of the \emph{chemical master equation} (CME):
\begin{equation}
\frac{\partial}{\partial t}p(t,x) = \sum_{j=1}^M \big(a_j(x-v_m)p(t,x-v_m)- a_j(x)p(t,x)\big). \label{eqn:chemical-master-equation}
\end{equation}
This solution is only analytically tractable in the simplest of models. In more complicated models with discretely observed data, standard statistical methods for performing inference on the $\theta_j$s are unavailable due to intractability of this probability distribution. This necessitates the use of analytical or numerical approximations such as approximate Bayesian computation \citep{Marj:Moli:Plag:Tava:mark:2003}.

\subsection{Stochastic simulation algorithm \label{sec:ssa}}

The DA-MCMC algorithm described later requires forward simulation of the system from a known initial state which is accomplished using the stochastic simulation algorithm (SSA) \citep{Gill:exac:1977}.  The basis of this algorithm is the \emph{next reaction density function} \citep{Gill:appr:2001}: 
\begin{equation}
p(\tau,J=j|x_t) = \frac{a_j(x_t)}{a_0(x_t)}\cdot a_0(x_t) e^{-a_0(x_t)\tau} \nonumber 
\end{equation}
where $a_0(x_t) = \sum_{j=1}^M a_j(x_t)$. Since the joint distribution is the \emph{product} of the marginal probability mass function for the reaction indicator $j$ and the probability density function for the next reaction time $\tau$, the reaction indicator and reaction time are independent. SSA involves sampling the reaction indicator $J=j$ with probability $a_j(x_t)/a_0(x_t)$ and the reaction time $\tau\sim Exp(a_0(x_t))$, where $Exp(\lambda)$ is an exponential random variable with mean $1/\lambda$. The state of the system is incremented according to the state change vector $v_j$, time is incremented by $\tau$, and the propensities are recalculated with the new system state. This process continues until the desired ending time is reached. Many speedups/approximations for SSA are available and the reader is referred to \cite{Gill:stoc:2007} for a review.

\section{Bayesian inference \label{sec:inference}}

Bayesian inference is a methodology that describes all uncertainty through probability. Let $y$ denote any data observed from the system and $\theta=(\theta_1,\ldots,\theta_M)'$ the vector of unknown parameters. The objective of a Bayesian analysis is the \emph{posterior distribution} that can be found using Bayes' rule
\begin{equation}
p(\theta|y) = \frac{p(y|\theta)p(\theta)}{p(y)} \propto p(y|\theta)p(\theta) \label{eqn:bayes-rule}
\end{equation}
where $p(y|\theta)$ is the statistical model, often referred to as the likelihood, $p(\theta)$ is the \emph{prior distribution} encoding information about the parameters available prior to the current experiment, and $p(y)$ is the normalizing constant to make $p(\theta|y)$ a valid probability distribution. The second half of equation \eqref{eqn:bayes-rule} indicates that it is rarely necessary to determine the normalizing constant $p(y)$ when performing a Bayesian inference.

\subsection{Complete observations \label{sec:complete-observations}}

In the unrealistic setting where the system is observed completely, i.e. $y=X=X_{[0,T]}$ where $X_{[a,b]}$ indicates all values for $X$ on the interval $[a,b]$, stochastic reaction rates can be inferred easily. If we assume independent gamma priors for each of the stochastic reaction rates, i.e. $\theta_j\stackrel{ind}{\sim} Ga(\alpha_j,\beta_j)$, where the gamma distribution is proportional to $\theta_j^{\alpha_j-1}\exp(-\theta_j\beta_j)$, then the posterior distribution under complete observations are independent gamma distributions
\begin{equation}
\theta_j|y \stackrel{ind}{\sim} Ga\left(\alpha_j+r_j, \beta_j+b_j \right) \label{eqn:theta-posterior}
\end{equation}
where $r_j$ is the number of times reaction $j$ fired and $b_j$ is the integral of $h_j(\cdot)$ over interval $[0,T]$.  Mathematically, we write
\begin{equation}
\begin{split}
r_j &= \sum_{k=1}^K \mathrm{I}(j_k=j) \\
b_j &= \int_0^T h_j(X_t) dt = \sum_{k=1}^K h_j(X_{t_{k-1}})\left(t_{k}-t_{k-1}\right) \label{eqn:sufficient-statistics}
\end{split}
\end{equation}
where $j_k\in\{1,\ldots,M\}$ is the reaction index for the $k$\textsuperscript{th} reaction, $\mathrm{I}(x)$ is the indicator function that is 1 when $x$ is true and 0 otherwise, $t_{k}$ is the time of the $k$\textsuperscript{th} reaction, and a total of $K$ reactions fired in the interval $[0,T]$. The two values $r_j$ and $b_j$ are the sufficient statistics for parameter $\theta_j$ and are utilized in Section \ref{sec:sufficient-statistics} to increase computational efficiency.

\subsection{Discrete observations}

In the more realistic scenario of perfect but discrete observations of the system, the problem becomes analytically intractable and, even worse, numerical techniques are challenging \citep{Wilk:stoc:2006}. This challenge comes from the necessity to simulate paths from $X_{t_{i-1}}$ to $X_{t_{i}}$ where $t_{i-1}$ and $t_{i}$ are consecutive observation times \citep{Boys:Wilk:Kirk:baye:2008}. These simulated paths are necessary when using a Gibbs sampling approach presented in the following section that alternates between 1) draws of parameters conditional on a trajectory and 2) draws of trajectories consistent with the data and conditional on parameters.

We define the discrete observations as $y=\{X_{t_i}: i=0,\ldots,n\}$ and update Bayes' rule in equation \eqref{eqn:bayes-rule} to include the unknown full latent trajectories $X$. The desired posterior distribution is now the joint distribution for the underlying latent states and the unknown parameters
\begin{align*}
p(\theta,X|y)  &\propto 
\mathrm{I}(y_0=X_{t_0})p(\theta)\prod_{i=1}^n \mathrm{I}(y_i=X_{t_i}) p\left(X_{(t_{i-1},t_i]}|\theta,X_{t_{i-1}}\right)  \notag \\
&=p(\theta)\prod_{i=1}^n p\left(X_{(t_{i-1},t_i)}|\theta,X_{t_{i-1}}=y_{i-1}, X_{t_i}=y_i\right)  \notag
\end{align*}
where 
 $p\left(X_{(t_{i-1},t_i)}|\theta,X_{t_{i-1}}=y_{i-1}, X_{t_i}=y_i\right)$ denotes the distribution for the latent state starting at $y_{i-1}$ and ending at $y_i$ over the time interval $(t_{i-1},t_i)$, i.e. the distribution for a continuous-time Markov chain with fixed endpoints. Since this distribution is not analytic -- even up to a proportionality constant -- the distribution $p(\theta,X|y)$ is not analytic.

\subsection{Markov chain Monte Carlo \label{sec:mcmc}}

A widely used technique to overcome these intractabilities in Bayesian analysis is a tool called Markov chain Monte Carlo (MCMC). In particular, we utilize a special case of MCMC known as Gibbs sampling  \citep{Marj:Moli:Plag:Tava:mark:2003}. This iterative approach consists of two steps:
\begin{enumerate}
\item draw $\theta^{(i)} \sim p(\theta|X^{(i-1)},y)$ and
\item draw $X^{(i)}\sim p(X|\theta^{(i)},y)$
\end{enumerate}
where superscript $(i)$ indicates that we use the draw from the $i$\textsuperscript{th} iteration of the MCMC, $p(\theta|X^{(i-1)},y)$ is the distribution for the parameters based on complete trajectories, and $p(X|\theta^{(i)},y)$ is the distribution for the complete trajectory based on the current parameters. The joint draw $\left(\theta^{(i)},X^{(i)}\right)$ defines an ergodic Markov chain with stationary distribution $p(\theta,X|y)$. 

In order to utilize this Gibbs sampling approach, samples from the \emph{full conditional distributions} for $\theta$ and $X$ are required, i.e. $p(\theta|X,y)$ and $p(X|\theta,y)$, respectively. Recognize that $p(\theta|X,y)=p(\theta|X)$ since $y\!\subset\! X$ due to $X$ representing the entire latent trajectory at all time points as if we had complete observations. Under complete observations, this full conditional distribution was already provided in equation \eqref{eqn:theta-posterior}. Therefore, samples are obtained by calculating the sufficient statistics in equation \eqref{eqn:sufficient-statistics} based on $X$ rather than $y$ and drawing independent gamma random variables for each reaction parameter.

\subsubsection{Rejection sampling \label{sec:rejection-sampling}}

The full conditional distribution for $X$ is, again, analytically intractable, but it is still possible to obtain samples from the distribution using \emph{rejection sampling} \cite[Ch. 2]{Robe:Case:mont:2004}. To accomplish this, we use Bayes' rule on the full conditional for $X$:
\begin{equation}
p(X|\theta,y) \propto \prod_{i=1}^n p\left(X_{(t_{i-1},t_i)}|\theta,X_{t_{i-1}}=y_{i-1}, X_{t_i}=y_i\right)  \notag
\end{equation}
where $p(\theta)$ is subsumed in the proportionality constant. A rejection sampling approach consistent with this full conditional distribution can be performed independently for each interval $(t_{i-1},t_i)$ for $i=1,\ldots,n$ as follows
\begin{enumerate}
\item forward simulate $X$ on the interval $(t_{i-1},t_i)$ via SSA using parameters $\theta$ with initial value $X_{t_{i-1}}=y_{i-1}$, and
\item accept the simulation if $X_{t_i}$ is equal to $y_i$ otherwise return to step 1.
\end{enumerate}
As with any rejection sampling approach, the efficiency of this method is determined by the probability of forward simulation from $X_{t_{i-1}}\!=y_{i-1}$ using parameters $\theta$ resulting in $X_{t_i}$ at time $t_i$. For these stochastic kinetic models, this probability can be very low and therefore we aim to take advantage of the massively parallel nature of graphical processing units to forward simulate many trajectories in parallel until one trajectory succeeds.

\section{Adaptation to graphical processing units \label{sec:gpu-adaption}}

Parallel processing is clearly only advantageous if the code is parallelizable. In the DA-MCMC algorithm, each step of the algorithm given in Section \ref{sec:mcmc} can be parallelized. The first step involves a joint draw for all stochastic reaction rate parameters conditional on a simulated trajectory. These parameters can be sampled independently as shown in equation \eqref{eqn:theta-posterior}. We can conceptually send the sufficient statistics for each reaction off to its own thread to sample a new rate parameter for that reaction. The second step involves rejection sampling to find a trajectory that satisfies interval-endpoint conditions. Both of these steps are candidates for parallel processing.

The theoretical maximum speed-up for parallel processing versus serial processing is bounded by Amdahl's quantity $1/(1-P+P/C)$ where $P$ is the percentage of time spent in code that can be parallelized and $C$ is the number of parallel cores available \citep{Amda:vali:1967, Such:Wang:Chan:Frel:Cron:West:unde:2010}. The first step in the DA-MCMC can only be parallelized up to the minimum of $M$, the number of reactions, and $C$. With the chemical kinetic systems under consideration, $M<<C$ and therefore the gain from parallelizing this step is minimal. In contrast, the gain in parallelizing the rejection sampling step in DA-MCMC is entirely dependent on the acceptance rate. As the acceptance rate drops, $P\to 1$ and maximum performance gain from parallelization is achieved. In our application, the acceptance rate is often very low and therefore we focus on efficiency gained from parallelizing the rejection sampling step.

Consider initially the goal of sampling a trajectory from some initial state $x_0$ at time 0 given parameter $\theta$ to a final state $x_1$ in time 1. Using SSA, the probability of simulating this trajectory by starting at $x_0$ using parameter $\theta$ and attaining $x_1$ has probability $p$, which is generally unknown. The number of simulations required before a successful attempt is a geometric random variable with expectation $1/p$. Therefore as the probability decreases, the expected number of runs increases and the system becomes amenable to parallel processing.

A simple approach to parallelization is to have each computing \emph{thread} attempt one simulation and determine whether that simulation was successful, i.e. the final state is $x_1$ at time 1. If multiple threads are successful, then one of those simulations is sampled uniformly. The sufficient statistics for that simulation are calculated and the DA-MCMC can continue by sampling rate parameters based on those statistics. In the following subsections, we discuss the implementation details required to turn this simple idea into a an efficient reality.

\subsection{Independent pseudo-random number streams}

Most current GPU-parallelized Monte Carlo algorithms  know, prior to parallel kernel invocation, how many random numbers will be needed by each thread and can therefore use a ``skip-ahead'' technique \citep{Lecu:Sima:Chen:Kelt:obje:2002} for obtaining independent streams of pseudo-random numbers \citep{Lee:Yau:Gile:Douc:Holm:on:2009}. This technique relies on one long pseudo-random number stream. The key idea is to have the next thread skip over the $n$ random numbers needed by the thread before it. Unfortunately, in the SSA algorithm the number of random numbers required for each thread is random and therefore this approach becomes infeasible unless we are willing to settle for $n$ sufficiently large to ensure a minuscule probability of overlap in the sequences. Even then, we are inviting computational inefficiency since the ``skip-ahead' requires $O(\log n)$ operations \citep{Lecu:Sima:Chen:Kelt:obje:2002}.

Instead, we use dynamic creation of pseudo-random numbers \cite{Mats:Nish:dyna:2000} which has already been used in the SSA context \cite{Li:Petz:effi:2010}. This method creates a set of pseudo-random number streams based on the Mersenne-Twister family of generators. A hypothesis concerning statistical independence of these streams is given in \cite{Mats:Nish:dyna:2000}: `a set of PRNGs [pseudo-random number generators] based on linear recurrences is mutually ``independent'' if the characteristic polynomials are relatively prime to each other.' These authors state that there are many PRNG researchers who agree with this hypothesis. 

\begin{figure}[htbp]
\centering
\includegraphics[width=5in,trim=0 2in 0 0, clip]{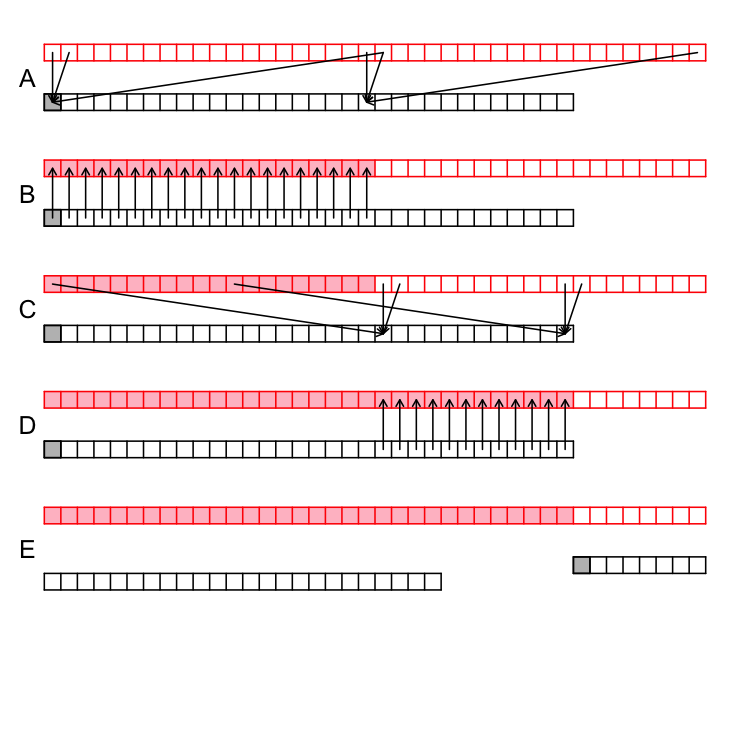}
\caption{A depiction of threads (black boxes, thread 1 is greyed) within a warp accessing Mersenne-Twister states (red boxes). A) The first 20 threads (threads 1 and 20 depicted) accessing states $i$, $i+1$ mod 40, and $i+20$ mod 40 where $i$ is the tread id. B) These threads update their corresponding Mersenne-Twister state (filled red boxes). C-D) Threads 21 to 32 now perform steps A and B. E) For the next round of pseudo-random numbers, the threads shift so that thread 1 starts at the next Mersenne-Twister state to be updated.}
\label{fig:prng-warp}
\end{figure}

To balance memory constraints (discussed in Section \ref{sec:mem-usage}) with execution speed, we implement one ``independent'' Mersenne-Twister (MT) per \emph{warp}, the set of threads that receive instructions simultaneously. Current generation GPUs have 32 threads per warp and, again balancing memory with execution, we have implemented MTs that utilize a set of 40 integers for calculation of pseudo-random numbers. Figure \ref{fig:prng-warp} depicts threads within a warp accessing one MT.  Within a warp, the first 20 threads execute simultaneously followed by the remaining 12 in order to ensure proper updating of the MT. To update the MT, thread $i$ records the MT state integers at locations $i$, $i+1$ modulo (mod) 40, $i+20$ mod 40 (operations specific to this 40-integer MT) \cite{Mats:Nish:dyna:2000}. The thread performs bit-wise operations on these integers and records the output as the updated state at location $i$ \cite{Mats:Nish:mers:1998}. After all threads have completed, they compute the pseudo-random number required for the SSA algorithm. For the following round of pseudo-random numbers, the threads are shifted with respect to the MT state such that thread $i$ is now aligned with MT state $i$+32 mod 40, a process that continues \emph{ad infinitum}. After all threads within a warp have completed the SSA algorithm, the MT state plus the last used state index are written to global memory for use in the following kernel invocation. 

\subsection{Bypass thread simulation}

Ideally once one thread is successful, current and future threads should be aborted. One aspect of GPU computing that varies from standard parallel processing is that no assumption can be made about the order in which threads occur. A statement such as `if all threads with global thread id less than $i$ have failed, then perform simulation' cannot be made since it is unknown which threads have already made an attempt. Nonetheless, the same effect can arise by creating a global variable that indicates when a thread has been successful, but care is required.

At any given time during the parallel execution of rejection sampling, threads can be placed into three categories: \emph{already-completed-and-failed}, \emph{in-progress}, and \emph{waiting-to-be-executed}. Once an \emph{in-progress} thread is successful, it writes to global memory indicating that it was successful and stores its pseudo-random number state. For \emph{already-completed-and-failed} threads no efficiency gains are possible. For \emph{waiting-to-be-executed} threads a simple check at the onset of simulation to the global success variable is sufficient to bypass the thread simulation if a thread has already been successful. Finally, \emph{in-progress} threads could periodically check whether another thread has been successful, but this adds unnecessary overhead and, more importantly, could bias results. Instead, \emph{in-progress} threads are allowed to complete even if another thread has been successful in the meantime. In the unlikely event that another thread is successful, it is added to the list of successful threads and at the completion of all threads one of the successful threads is sampled uniformly.

\subsection{Sufficient statistics \label{sec:sufficient-statistics}}

A naive approach to recording the SSA trajectory is to record the state and time when each reaction fires which requires $NK+K$ integers/foats where $K$ is the number of reactions that fired. A parsimonious way of representing a trajectory is to record the initial state vector as well as the times and identity of each reaction and recreating the trajectory when needed. The memory storage requirement for this approach is $N+2K$ integers/floats. In addition, $K$ is random and therefore memory management is required to deal with varying array sizes.

Recall that the full condition distribution for the rate parameters depends only on the sufficient statistics in equation \eqref{eqn:sufficient-statistics}. So rather than recording the entire trajectory, the sufficient statistics can be recorded which reduces the memory requirement down to $2M$ integers/floats where, often, $M\approx N$ and $K>>M$. If inference on the trajectories is required, then this can be accomplished after the DA-MCMC is complete by rerunning the rejection sampling for each (or a subset) of the MCMC iteration values for the rate parameters.

A solution that reduces the memory requirements even further and does not require rerunning the rejection sampling is to record the initial PRNG state that resulted in a successful simulation in lieu of both the full trajectory and the sufficient statistics.  The memory storage requirement is only 41 integers (40-integer MT state plus the last used state index) which clearly scales with $N$, $M$, and $K$. For calculation of sufficient statistics or any trajectory inference, the trajectory is re-simulated with parameters corresponding to that iteration in the MCMC and using the initial PRNG state that was previously successful.

\subsection{Efficient memory usage \label{sec:mem-usage}}

\begin{figure}[tbp]
\centering
\includegraphics[width=3in]{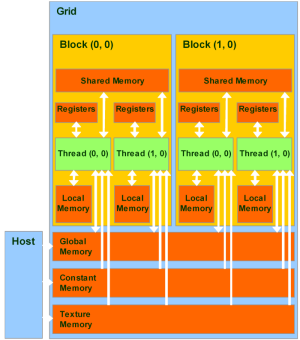}
\caption{The CUDA memory model.}
\label{fig:memory-model}
\end{figure}

A trade-off made for the massively parallel nature of the GPU is the amount of memory available for each thread. This is an overarching concern that has been covered elsewhere \citep{Lee:Yau:Gile:Douc:Holm:on:2009, Such:Wang:Chan:Frel:Cron:West:unde:2010}, but we now discuss how this concern can be addressed in the parallel rejection sampling framework. Figure \ref{fig:memory-model} provides a diagram of the CUDA memory model. Importantly, any access to memory below the threads (in green) is relatively slow and memory depicted above the threads is fast, but very limited.

\begin{table}
\begin{minipage}{\textwidth}
\caption{Relevant hardware memory constraints in kilobits (integers) for the Tesla T10 GPU and algorithm memory allocation.}
\label{tab:t10-memory}
\end{minipage}

\vspace{0.1in}

\centering
\begin{tabular}{|l|rlc|l|}
\hline
Memory type & \multicolumn{3}{c|}{Amount (integers)} & Algorithm allocation \\
\hline
Registers per block & \multicolumn{2}{c}{$32\times 512$} && Thread system time \\
&&&& Thread loop variables \\
Shared per SM & 4 & kb &($8\times512$) & Twister state during SSA simulation \\
&&&&  Thread system state $\dagger$ \\
&&&& Thread reaction propensities $\dagger$ \\
\hline
Local per thread & 16 & kb & (4096) & Thread system state $\dagger$ \\
&&&& Thread reaction propensities $\dagger$ \\
Global & 4 & Gb &($\approx 10^9$) & Success counter \\
&&&& Successful twister state \\
&&&& Twister states when not in use \\
Constant & 64 & kb &(16,384) & Reaction rate parameters \\
&&&& Stoichiometry matrix \\
\hline
\multicolumn{5}{l}{$\dagger$ If shared memory is available, thread system state and reaction propensities} \\
\multicolumn{5}{l}{$\phantom{\dagger}$ are moved from local to shared memory.}
\end{tabular}
\end{table}

Table \ref{tab:t10-memory} provides hardware memory constraints on the Tesla T10 GPU and the algorithm memory allocation described here. Constant memory can be used for quantities that do not change during the GPU kernel execution and has an 8k cache for efficient access \cite[Ch. 5]{Kirk:Hwu:prog:2009}. Therefore it is a convenient location for the stoichiometry matrix and reaction rate parameters (which only change outside of the GPU kernel) since these are common to all blocks and threads. Without considering efficient sparse matrix storage, the number of integers needed to store both the matrix and parameters is $M(N+1)$ which, given the systems of interest, easily fits within the constant memory cache.

After constant memory, efficient memory access is achieved by utilizing registers and shared memory. Registers are restricted to automatic scalar variables, i.e. not arrays, that are unique to each thread \cite[Ch. 5]{Kirk:Hwu:prog:2009}. If we are using the maximum number of threads per block of 512, then we are limited to 32 registers. Clear candidates for register storage are system time in the SSA simulation and simple loop variables. To determine the number of registers per thread compile using the nvcc compiler option \texttt{--ptxas-options=-v}. The SSA algorithm currently uses all 32 registers per thread and therefore allows us to use the maximum of 512 threads per block. 

Shared memory is fast-access memory that can be utilized by all threads within a block. In this implementation, we take advantage of the fast-access by storing pseudo random number generator states.  Also, depending on the system size, we can store each thread's SSA current system state, an $N$-integer array, and possibly even the $M$-float array containing the reaction propensities.  The PRNGs take up 2560 bytes of shared memory per block leaving 13824 bytes left over to store the system state and/or the reaction propensities.  Therefore, if $N + M$ $\leq$ $6$, then both the state and propensities can be stored, otherwise there is not enough memory available to store both.  This limitation can be met by either decreasing the number of threads per block to increase the amount of available shared memory per thread or by using local or global memory.  In our experience, the optimal method generally depends on the system being studied.

Remaining variables are stored either in local or global memory depending on whether they are thread-specific or not. For example, two important global variables include the indicator of whether a thread has been successful and the array of successful MT states.

\section{Simulation example \label{sec:simulation}}

We compare the efficiency of a GPU implementation to conventional CPU implementation using the model Michaelis-Menton model. This widely known model is described by the following reaction graph:
\begin{equation}
E + S \xrightleftharpoons[\theta_2]{\theta_1} ES \stackrel{\theta_3}{\longrightarrow} E + P \label{eqn:michaelis-menton}
\end{equation}
where $E$ is a protein enzyme, $S$ is a substrate that is converted into a product $P$, and $ES$ is an intermediate species for this production. The propensities for the three reactions are $a_1(X)=\theta_1E\cdot S$, $a_2(X)=\theta_2ES$, and $a_3(X)=\theta_3ES$ where $X=(E,S,ES,P)$. This system has two conservation of mass relationships: $E_0=E+ES$ and $S_0=S+ES+P$.

\subsection{GPU vs CPU}

For comparing GPU vs CPU timing, we considered only the rejection sampling step in Section \ref{sec:rejection-sampling} of the DA-MCMC algorithm rather than the entire MCMC algorithm. This was done since the probability of rejection is highly dependent on the current MCMC parameter draws and to obtain accurate timing a large quantity of MCMC iterations is required to eliminated timing bias due to exploration of the parameter posterior. Unfortunately, obtaining a reasonable quantity of MCMC iterations on the CPU is simply not feasible on a reasonable time scale. Therefore, we compare timing for the rejection sampling portion of the algorithm only, although we expect the efficiency gain for the GPU should be comparable if the entire MCMC timing could be analyzed.


\begin{figure}[htb]
\begin{center}
\includegraphics[width=5in]{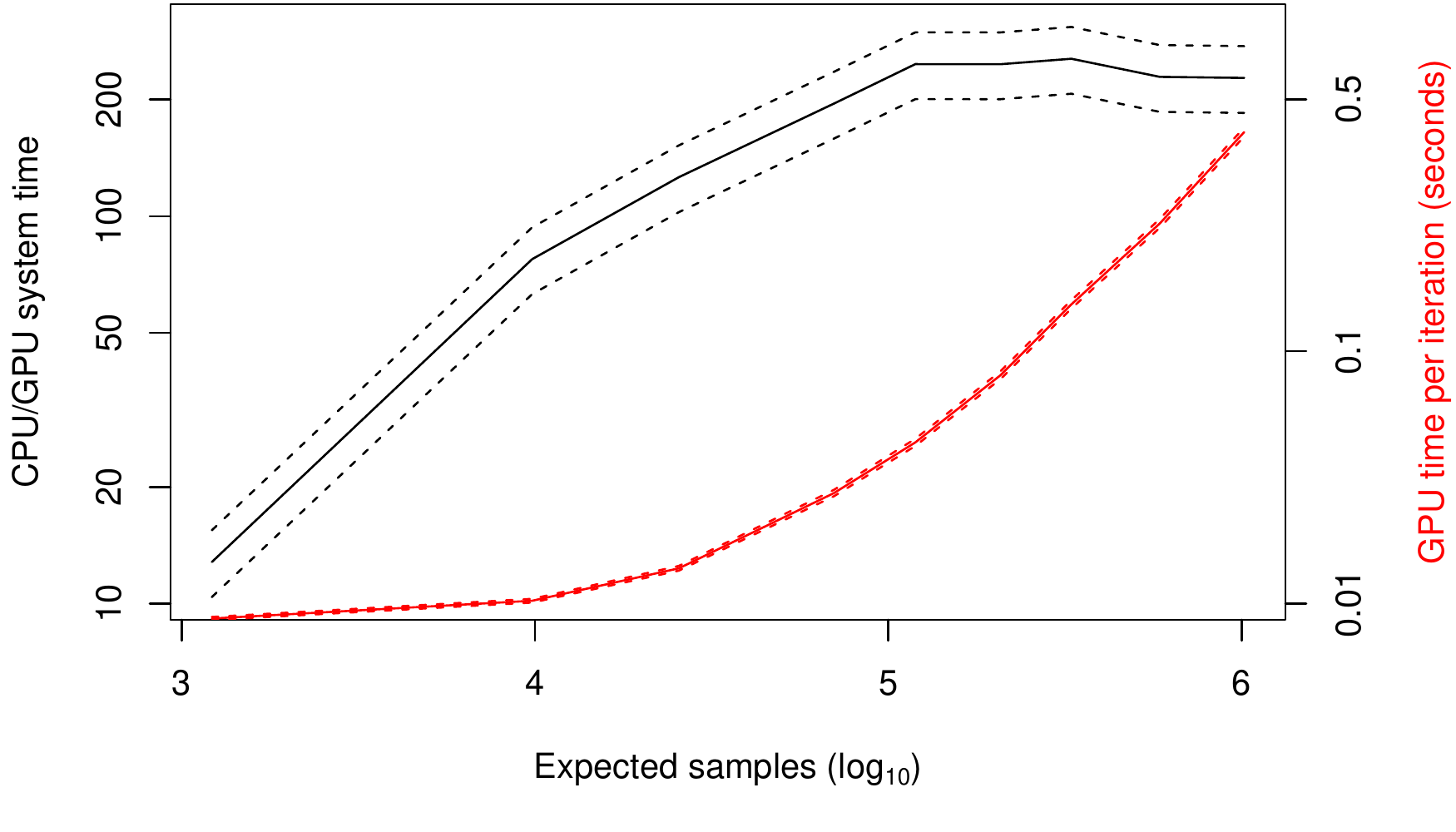}
\caption{The multiplicative increase in rejection sampling efficiency (black) and the GPU time for one iteration (red) with 95\% intervals (dashed) for one core of a 2.66GHz Intel Xeon CPU vs a Tesla T10 GPU. The effect of expected number of samples required for each acceptance was studied by holding constant $\theta_2=0.001$ and $\theta_3=0.1$ while increasing $\theta_1$ from 0.001 to 0.00245 in the Michaelis-Menton system of equation \ref{eqn:michaelis-menton}.}
\label{fig:cpu-gpu-timing}
\end{center}
\end{figure}

It is important to note that if the rejection sampling step had no rejections, then we would expect the CPU to perform comparable to the GPU and possibly even better if GPU overhead is considerable. Therefore it is of interest to study the efficiency gain as a function of the difficulty of the rejection sampling step. Figure \ref{fig:cpu-gpu-timing} compares the efficiency of one core of a 2.66GHz Intel Xeon CPU vs one Tesla T10 GPU where increasing difficulty of rejection sampling is equivalent to increased expected draws. 
For high acceptance probability rejection sampling schemes, the efficiency gain is modest and may not be worth the trouble of converting code to GPU use. In contrast for low acceptance probability rejection sampling schemes, the efficiency gain is around 200, meaning the GPU version will perform 200 times faster than the CPU version. The efficiency gain appears to hit an asymptote around 200, for our algorithm implementation while the computation time involved appears to be exponentially increasing as the acceptance probability decreases. Therefore incredibly low acceptance probability rejection sampling schemes could still not be handled on a GPU, but may be suitable to simultaneous use of multiple GPUs.

\subsection{Bayesian inference}

The ultimate goal of this Bayesian analysis is performing inference in stochastic chemical kinetic models on both the unknown reaction rate parameters and the latent trajectory between data points. The Michaelis-Mention system was simulated with true parameters $\theta_1=0.001$, $\theta_2=0.2$, and $\theta_3=0.1$ from time 0 up to time 100. The data used are provided in Table \ref{tab:data} where both the $ES$-complex and product $P$ are initialyl zero. Through the monotonic decrease in $S$, it is clear this system converts the substrate in the product. The enzyme quantity initially decreases drastically as it bonds to available substrate, but then as substrate is converted to product more unbound enzyme is available.
\begin{table}[htb]
\begin{center}
\begin{minipage}{0.85\textwidth}
\caption{Measurements taken from a simulated Michaelis-Mention system with parameters $\theta_1=0.001$, $\theta_2=0.2$, and $\theta_3=0.1$.}
\label{tab:data}
\end{minipage}
\begin{tabular}{|l|rrrrrrrrrrr|}
\hline
Time & 0 & 10 & 20 & 30 & 40 & 50 & 60 & 70 & 80 & 90 & 100 \\
\hline
$E$ & 120 & 71 & 76 & 81 & 80 & 90 & 90& 104 & 103 & 109 & 109 \\
$S$ & 301 & 219 & 180 & 150 & 108 & 86 & 61 & 52 & 35 & 29 & 22 \\
\hline
\end{tabular}
\end{center}
\end{table}

A non-informative independent prior is assumed for all reaction rate parameters, namely $p(\theta)\propto (\theta_1\theta_2\theta_3)^{-1}$. This prior is found as the limit of the gamma priors when both the shape and rate parameters approach zero, but the gamma posterior of equation \eqref{eqn:theta-posterior} is still proper with $\alpha_j=\beta_j=0\forall j$ if each reaction occurs at least once.

The DA-MCMC algorithm was run for 10,000 burn-in iterations and then another 40,000 iterations were used for inference. Convergence was monitored informally via traceplots and formally using the Gelman-Rubin diagnostic \citep{Gelm:Rubi:infe:1992,Broo:Gelm:gene:1997}. While lack of convergence is detected, samples are discard as \emph{burn-in}. Post burn-in, no lack of convergence was detected. 

\begin{figure}[htb]
\begin{center}
\includegraphics[width=5in]{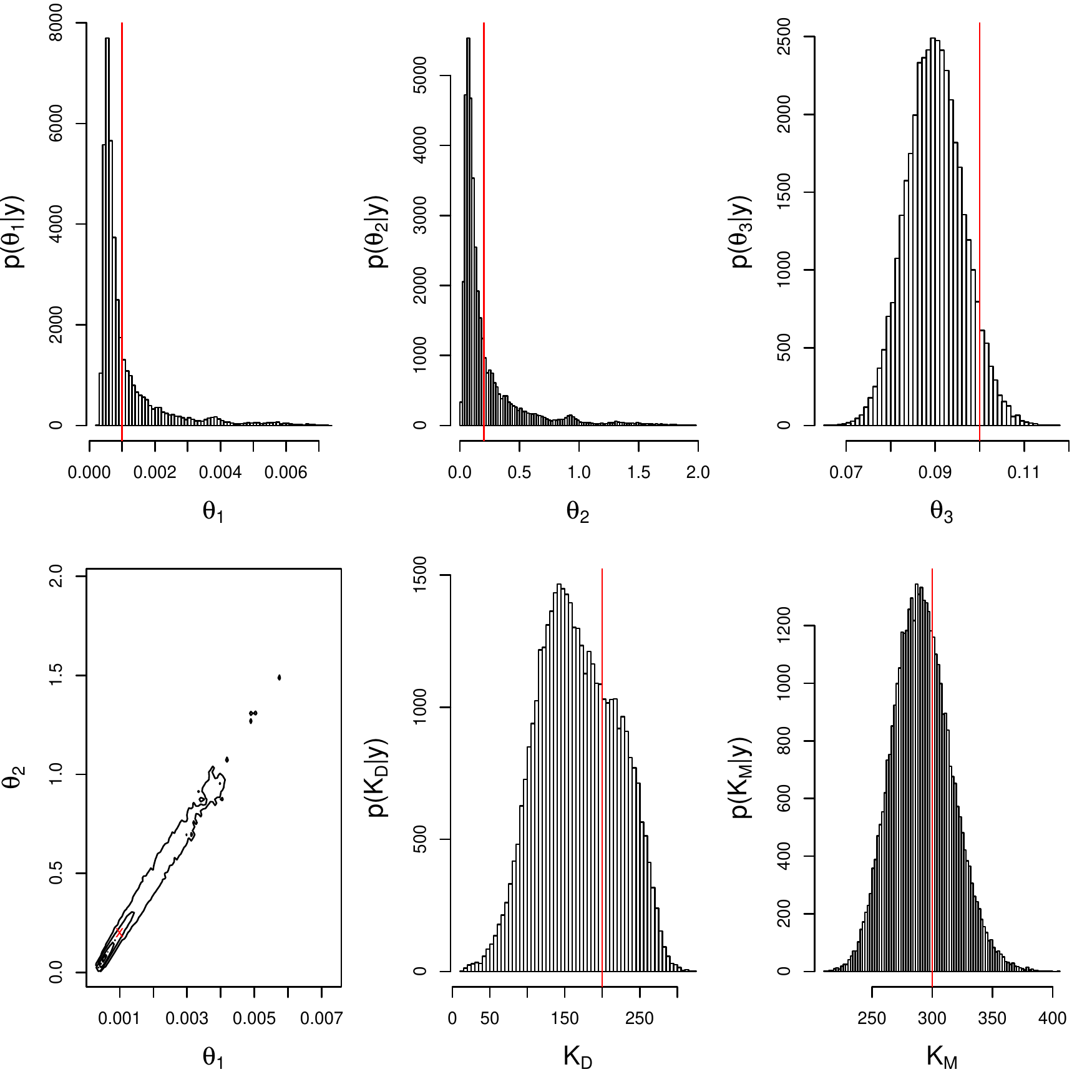}
\caption{Posterior histograms for stochastic reaction rate parameters as well as the stochastic dissociation and Michaelis constants, $K_D=\theta_2/\theta_1$ and $K_M=[\theta_2+\theta_3]/\theta_1$ respectively, and a bivariate contour plot (quantiles: 2.5\%, 25\%, 50\%, 75\%, and 95\%) for the joint posterior of $\theta_1$ and $\theta_2$ with true values (red) based on the data in Table \ref{tab:data} and using the DA-MCMC algorithm. }
\label{fig:params}
\end{center}
\end{figure}

Figure \ref{fig:params} provides posterior histograms for the reaction rate parameters based on the observations in Table \ref{tab:data}.
The bivariate contour plot of $\theta_1$ and $\theta_2$ indicate that the value $\theta_1+\theta_2$ is estimable from the data, but the individual values for $\theta_1$ and $\theta_2$ are hard to estimate. This identifiability issue is common in systems biological parameter inference where equilibrium reactions abound. 

One advantage of Bayesian analyses is trivially obtained estimates and uncertainties for any function of the model parameters. Figure \ref{fig:params} provides posterior histograms for both $K_D=\theta_2/\theta_1$ and $K_M=[\theta_2+\theta_3]/\theta_1$ known as the dissociation constant and Michaelis constant, respectively. Although many methods have been developed to estimate the Michaelis constant, dating at least to the Lineweaver-Burk plot from 1934 \cite{Line:Burk:dete:1934}, few methods provide an uncertainty on the estimate. Based on the plot in Figure \ref{fig:params}, there is 95\% probability that the true is in the range 246 to 343. 

\begin{figure}[htbp]
\begin{center}
\includegraphics[width=5in]{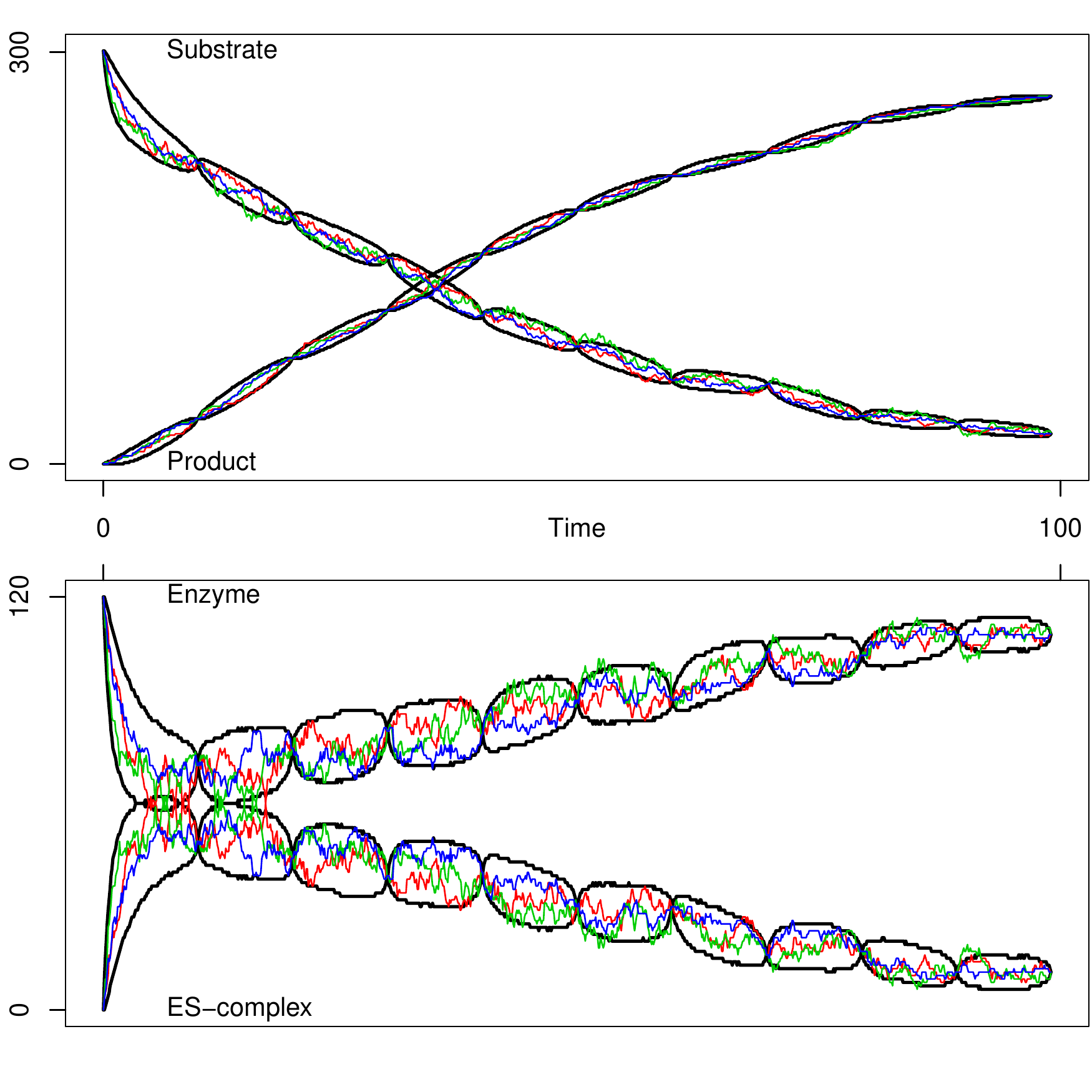}
\caption{Posterior 95\% point-wise credible intervals (black) and three random draws from the posterior (colored) for the state trajectory.}
\label{fig:trajectories}
\end{center}
\end{figure}

Figure \ref{fig:trajectories} provides point-wise credible intervals for the four Michaelis-Menton system species.
Due to mass conservation, we see that $E$ and $ES$ are mirror opposites of each other and $S$ and $P$ are very close to being mirror opposites of each other. The scientific questions of interest that are answered by these trajectories include \emph{when was the $S\to P$ conversion 90\% complete?} or \emph{what is the probability that $ES$ crossed the 90 molecule threshold?} Bayesian analyses can trivially answer these questions while it remains difficult for other statistical methods.

\section{Discussion \label{sec:discussion}}

We presented a Bayesian analysis of stochastic chemical kinetic models that utilize a data augmented MCMC algorithm where the augmentation infers latent trajectories sampled via rejection sampling. This dramatic increase in efficiency when utilizing a GPU will allow for analysis of vastly larger systems in reasonable amounts of time. The timing comparison in this manuscript only compared rejection sampling and therefore the results are biased slightly in favor of the GPU. Further work is required to explore the efficiency gain of the entire MCMC, but we suspect the results to be very similar to the results presented here and the benefit of letting the CPU algorithm run for months is marginal.

The observations in this manuscript were discrete but perfect. Clearly this is an unrealistic scenario in practical applications since we rarely obtain perfect observations of the underlying system. Realistic models naturally incorporate error in one of two ways: 1) the true value is within a threshold of that observed or 2) all values are possible but values closer to that observed are more probable. In the first approach, everything discussed in this manuscript is still applicable since forward simulations that are consistent with the observations will still be needed. These simulations could easily be harder to obtain and therefore more amenable to parallelization. In the second approach, all trajectories are possibilities and therefore rejection sampling is not applicable. We are exploring the use of an independent Metropolis-Hastings proposal and  methodologies that exploit creation of multiple independent proposals simultaneously. 


Approximate Bayesian computation approaches have already been implemented on a GPU in a package called ABC-SysBIO \citep{Liep:Barn:Cule:Ergu:Krik:Toni:Stum:abc:2010}. This was implemented in Python utilizing the PyCUDA wrapper\citep{Kloc:Pint:Lee:Cata:Ivan:Fasi:pycu:2009} to access the CUDA API. Since \cite{Liep:Barn:Cule:Ergu:Krik:Toni:Stum:abc:2010} devotes only two paragraphs relevant to this manuscript, it is unclear how PyCUDA implements the algorithm, e.g. random number generation, memory efficiency, etc., and whether it is competitive with the implementation discussed in this manuscript. 

The few papers discussing Bayesian inference on GPUs published to date have shown remarkable efficiency gains. Since this field is computation heavy, this increased efficiency should lead to Bayesian techniques being much more widely adopted than they are today as the capacity to solve highly complex problems in reasonable time frames increases.

\section{Acknowledgements}
The authors gratefully acknowledge support from NSF IGERT Grant DGE-0221715 and the Institute for Collaborative Biotechnologies through contract no. W911NF-09-D-0001 from the U.S. Army Research Office.  The content of the information herein does not necessarily reflect the position or policy of the Government and no official endoresement should be inferred.

\bibliographystyle{plainnat}
\bibliography{gpgpu,jarad}

\end{document}